\documentclass[twocolumn]{aastex631}

\newcommand{\MgII}{\textsc{{\rm Mg}\kern 0.1em{\sc ii}}}
\newcommand{\CIV}{\textsc{{\rm C}\kern 0.1em{\sc iv}}}
\newcommand{\OVI}{\textsc{{\rm O}\kern 0.1em{\sc vi}}}
\newcommand{\OVII}{\textsc{{\rm O}\kern 0.1em{\sc vii}}}
\submitjournal{ApJ}
\graphicspath{{./}{figures/}}

\begin{document}

\title{The Thermodynamic and Kinematic Evolution of Circumgalactic Gas around $z=1$ in the IllustrisTNG model}

\correspondingauthor{Daniel DeFelippis}
\email{ddefelip@arizona.edu}

\author[0000-0002-0112-7690]{Daniel DeFelippis}
\affiliation{Department of Astronomy \& Steward Observatory, University of Arizona, 933 N. Cherry Avenue, Tucson, AZ 85721, USA}

\author[0000-0002-3185-1540]{Shy Genel}
\affiliation{Center for Computational Astrophysics, Flatiron Institute, 162 Fifth Avenue, New York, NY 10010, USA}
\affiliation{Columbia Astrophysics Laboratory, Columbia University, 550 West 120th Street, New York, NY 10027, USA}

\author[0000-0003-2630-9228]{Greg L. Bryan}
\affiliation{Columbia Astrophysics Laboratory, Columbia University, 550 West 120th Street, New York, NY 10027, USA}
\affiliation{Center for Computational Astrophysics, Flatiron Institute, 162 Fifth Avenue, New York, NY 10010, USA}

\begin{abstract}
The circumgalactic medium (CGM) is known to contain multiphase gas in various stages of evolution and interaction with the galaxy. In order to characterize its detailed behavior on short timescales, we use a subregion of the TNG100 cosmological simulation to study the evolution of the $z=1$ CGM around six galaxies in $10^{11.5}-10^{12}\,M_{\odot}$ halos at a high time cadence of $\approx2$ Myr. We use Monte Carlo tracer particles to follow this CGM gas forward in time in a Lagrangian way and determine how its thermodynamic and kinematic properties change. We find that CGM gas mixes between different temperature and density phases quickly and within $\approx500$ Myr evolves into distinct cold ($T\approx10^4\,\rm{K}$) and warm--hot ($T\approx10^{5.5}\,\rm{K}$) phases at small and large distances from the galaxy, respectively, regardless of its initial ($z=1$) halo-centric radius. This is largely driven by feedback from the galaxy, which heats and ejects cold gas that had previously cooled and accreted toward and occasionally into the galaxy from the outer CGM. We see signatures of this process in autocorrelations of kinematic quantities, which take $\approx400$ Myr to fully decorrelate from their initial values, suggesting a timescale over which feedback disrupts and reprocesses CGM gas. We also examine gas in narrow temperature and density ranges associated with commonly observed ions and find that gas that is \OVI--like stays in its phase for hundreds of Myr longer than gas that is \MgII--like or \CIV--like, suggesting that CGM observations of different species could probe gas in different evolutionary states, even if the gas is cospatial.
\end{abstract}

\keywords{Galaxy formation (595) --- Galaxy dynamics (591) --- Galaxy kinematics (602) --- Galaxy structure (622) --- Circumgalactic medium (1879) --- Hydrodynamical simulations (767)}

\section{Introduction} \label{sec:intro}

The circumgalactic medium (CGM) is a region surrounding galaxies that is composed of diffuse gas in a variety of different temperatures, densities, and kinematic states \citep[e.g.,][]{Tumlinson17,Faucher-Giguere23}. The CGM is believed to contain a combination of pristine gas that is accreting onto galaxies from the cosmic web and recycled gas that has been ejected from the galaxy. The way these diffuse halos of gas interact with galaxies is expected to influence the way galaxies grow and change over time, so characterizing and analyzing the state of the CGM is a crucial step toward building up a comprehensive picture of galaxy formation and evolution. 

The CGM has most commonly been observed using absorption-line spectroscopy, with background quasars providing sight lines that intersect circumgalactic regions of foreground galaxies. Surveys of galaxy--absorber pairs at lower redshifts such as COS-Halos \citep{Werk14} and those at intermediate and higher redshifts such as MAGIICAT \citep{Nielsen13a,Nielsen13b}, MEGAFLOW \citep{Bouche25}, KBSS \citep{Rudie12}, and CUBS \citep{Cooper21,Qu22,Qu23} have found that the CGM is an environment that extends significantly beyond the stellar extent of most galaxies. Furthermore, this environment is composed of gas of multiple temperatures and densities that are probed by different ions that themselves have distinct properties and distributions. For example, many surveys have focused on the \MgII{} ion \citep[e.g., MAGG;][]{Dutta20}, which traces colder and denser gas that tends to be found closer to the major and minor axes of the galaxy \citep{Bouche12,Nielsen15}. In particular, \MgII{} gas near the major axis of the galaxy is found to be both corotating with \citep{Zabl19} and potentially accreting onto \citep{Ho17} the galaxy. Other surveys like COS-Halos have helped establish that higher ions tend to be distributed more isotropically around galaxies \citep[][]{Werk16}, though still over a large range of velocities, and this hotter, more diffuse gas can still be rotating on large scales \citep{Hodges-Kluck16,Oppenheimer18}.

The CGM does not exist in isolation and is believed to be evolutionarily connected to its galaxy. By correlating quasar spectra from the Sloan Digital Sky Survey (SDSS) to galaxies in the DESI Survey, \cite{Lan20} was able to find connections between the galaxy and the ``cool'' ($\approx10^4 \, \rm{K}$) CGM observed with \MgII{} over cosmic time. In particular, they find that the covering fraction of the cool CGM is $\sim0.5$ dex higher around galaxies that are star-forming compared to those that are passive and that this relationship is consistent across stellar mass and redshift. \cite{Anand21} further make a distinction between the \MgII{} distributions around emission-line galaxies, where the covering fraction is most dependent on star-formation rate (SFR), and luminous red galaxies, where the covering fraction is most dependent on stellar mass. Similar results for the hotter \OVI{} phase are also found by \cite{Tchernyshyov23}: CGM covering fractions of star-forming galaxies are higher than those of passive galaxies over $\sim1$ dex in stellar mass for $z<0.6$. In simulations, \cite{Peroux20} find an angular dependence on the metallicity of CGM gas, with higher metallicities concentrated near the minor axis of galaxies, where feedback can more efficiently deposit enriched gas. Together, these results suggest that processes within the interstellar medium (ISM) of galaxies are influencing the development of the distribution and structures of gas observed in the CGM. Understanding this connection requires examining the time-series evolution of individual simulated galaxies. 

Generally, galaxy formation simulations focus their computational efforts on modeling processes within galaxies, including metal cooling, star formation, supernovae, and active galactic nuclei (AGN), while completely ignoring larger-scale environments like the CGM. These include simulations that model regions within and above/below galactic disks such as SILCC \citep{Walch15,Rathjen23}, TIGRESS \citep{Kim23}, and QED \citep{Vijayan24,Vijayan25}, as well as isolated galaxy simulations that expel mass into a uniform halo via outflows \citep{Sarkar15,Fielding17b,Fielding17a,Schneider18,Marinacci19,Shimizu19,Smith24}. Cosmological simulations like EAGLE \citep{Schaye15} and SIMBA \citep{Dave19}, which do explicitly evolve the CGM and larger scales, do so at comparatively low resolution owing to the CGM's diffuse nature. Nevertheless, studies have used these simulations to examine kinematic properties of the CGM at different epochs across cosmic time and have established potential connections between properties of the galaxy and the CGM \citep[e.g.,][]{Oppenheimer18a,Ho19,DeFelippis20}. Recently, there have also been studies that focused computational resources on resolving the CGM to a significantly higher resolution \citep[e.g.,][]{Hummels19,Peeples19,Suresh19,Ramesh24}. Results of these studies generally find more resolved structures, especially in the cold phase of the CGM, though the expected physical scales of these structures are not very well constrained \citep[e.g.,][]{Nelson20}. 

In addition to the recent work on characterizing and spatially resolving CGM gas in cosmological simulations, there have also been studies of the detailed time evolution of the CGM. Many recent analytic models of the CGM assume some kind of equilibrium state between multiple phases of gas \citep[e.g.,][]{Faerman17,Faerman20,Dutta24} and are fairly successful at reproducing CGM observations from absorption sight lines, especially for hotter gas, while models such as \textsc{cloudflex} \citep{Hummels24} and \emph{mCC} \citep{Bisht25} have focused on specifically modeling the denser cold phase in a turbulent medium and are also able to reproduce observational trends for ions like \MgII. Studies using a full-physics model such as \textsc{fire-2} \citep[][]{Hafen20} or \textsc{foggie} \citep[][]{Lochhaas23} find that the multiphase CGM is fed by many sources, does not often follow basic models like hydrostatic equilibrium, and varies significantly depending on the evolutionary stage of the galaxy. Furthermore, processes like gas mixing that are crucial in CGM evolution operate on much shorter timescales \citep[$\approx$a few to tens of Myr; see, e.g.,][]{Kwak10,Ji19a,Liang20,Fielding22,Shah25} than tend to be analyzed in simulations that model the CGM. Understanding how the CGM evolves over these shorter timescales is valuable for providing theoretical predictions for how both galaxies with different properties and simulation codes with different models of feedback \citep[][]{Fielding20,Strawn24} interact with and shape the CGM over time. 

In this paper, we present a Lagrangian evolutionary analysis of the intermediate-redshift ($z=1$) CGM of a small sample of simulated galaxies from the IllustrisTNG simulation suite. We use the high time cadence of the galaxy sample to study the evolution of the CGM around that redshift in great detail. The structure of the paper is as follows: In Section \ref{sec:methods} we highlight key aspects of the subset of TNG100 used in this paper and detail our analysis procedure. In Section \ref{sec:results} we describe our main results, first focusing on galaxy--CGM mixing timescales, then moving to the thermodynamic and kinematic evolution of the CGM, and finally connecting to ions used in quasar absorption-line studies of the CGM. In Section \ref{sec:discussion} we discuss possible physical drivers of the evolutionary trends, compare our results to other recent studies of the evolution of the CGM, and highlight how our results connect to observations of the CGM. We summarize our findings and conclude in Section \ref{sec:summary}.

\section{Methods} \label{sec:methods}

\subsection{Simulations} \label{subsec:simulations}

We utilize the TNG100 box from the IllustrisTNG simulation suite \citep{Marinacci18,Naiman18,Nelson18,Pillepich18,Springel18}. TNG100 uses the moving-mesh code \textsc{Arepo} \citep{Springel10,Weinberger20} to evolve a comoving box of size $\approx(111\, \rm{Mpc})^3$ to $z=0$ from cosmological initial conditions using cosmological parameters from \cite{Planck15}. Gas cells have a (baryonic) mass resolution of $\approx1.4\times10^6 \, M_{\odot}$. As detailed in \cite{Weinberger17} and \cite{Pillepich18a}, this simulation models star formation, chemical enrichment, metal cooling, magnetic fields, and feedback in the form of galactic winds launched by supernovae and energy injections from AGN powered by accretion onto black holes. We make use of the massless Monte Carlo tracer particles \citep{Genel13} stored in the simulation in order to follow the motion of gas as it moves between cells in the mesh, thus making a Lagrangian analysis of the mass in the CGM possible. To get the most evolutionary information, we focus on galaxy--CGM systems in TNG100 that are part of one of the subboxes, a region representing $0.1\%$ of the volume of the full box where there are 7908 saved snapshots (compared to 100 saved snapshots in the full box), representing a time cadence of $\sim2$ Myr (compared to $\sim140$ Myr for the full box). 

\subsection{Analysis} \label{subsec:analysis}

\begin{table*}
\centering
\begin{tabular}{lccccccr} \hline \hline
Halo ID & $M_{\rm{halo}}$ & $R_{\rm{vir}}$ & $M_*$ & $M_{\rm{CGM}}$ & $\overline{SFR}$ & $N_{\rm{tracers}}$ \\ 
& $ \log \, M_{\odot}$ & (kpc) & $ \log \,M_{\odot}$ & $ \log \, M_{\odot}$ & ($M_{\odot} \, \rm{yr}^{-1}$) \\ \hline
2371 & 11.94 & 149 & 9.98 & 11.01 & 4.05 & 137,925 \\  
3512 & 11.82 & 136 & 9.71 & 10.82 & 2.28 & 88,687 \\  
4316 & 11.71 & 124 & 9.83 & 10.67 & 1.47 & 65,269 \\ 
4530 & 11.71 & 125 & 9.73 & 10.66 & 0.97 & 62,219 \\ 
5360 & 11.61 & 116 & 9.27 & 10.60 & 1.29 & 57,084 \\ 
5819 & 11.55 & 110 & 8.99 & 10.62 & 0.41 & 58,242 \\ \hline 
\end{tabular}
\caption{Key properties of the six halos studied in this paper. The columns show the halo ID, halo mass, virial radius, galaxy stellar mass, CGM mass, SFR, and total number of tracers. All quantities are shown for $z=1$ except the SFR, which is averaged over the following 2 Gyr.}
\label{tab:halos}
\end{table*}

Our goal is to study a regime where galaxies are actively forming and the CGM is well observed \citep[e.g.,][]{Schroetter21}, so we focus our analysis at a redshift of $z\approx1$. We select halos at $z=1$ from the TNG100 fiducial sample of \cite{DeFelippis21} that are also present in TNG100-Subbox1 for at least 2 Gyr following that redshift: this results in six halos with halo masses at $z=1$ between $10^{11.5}$ and $10^{12} \, M_{\odot}$ ($0.14\%$ of the 4315 halos in this mass range in the full TNG100 box), where the mass of the halo is defined as in \cite{Bryan98}. We show key properties of our sample of six halos in Table \ref{tab:halos}. Of the multiple subboxes available from the TNG simulation suite, we choose to use TNG100-Subbox1 because, despite having a lower resolution than TNG50, it has the largest number of snapshots (and thus the best time cadence), more than one halo in our mass range of interest, and it does not contain any very massive halos that could significantly disrupt the evolution of the smaller halos we want to examine. Within each halo, we define the CGM of its central galaxy as all gas cells in the halo that are outside twice the stellar half-mass radius of the central subhalo, excluding any gas cells that are star-forming. We then identify all tracer particles associated with the CGM gas cells at $z=1$, locate those tracers in each subsequent snapshot for the following $\sim2$ Gyr (up to $z=0.6$, or $\sim 1$ dynamical time, where $t_{\rm{dyn}} = 2R_{\rm{vir}}/V_{\rm{vir}}$), and calculate desired thermodynamic and kinematic quantities of the gas associated with those tracers.\footnote{For a similar analysis using tracer particles in TNG50 focused on turbulence in the disk, see \cite{Forbes23}.} If a tracer enters a different physical state (star-forming gas, star, or black hole), we exclude it from our thermodynamic and kinematic analyses from that point forward until it rejoins the gas phase at a later time. For all six halos, we find but do not show that, at any one time, at most $10\%$, $5\%$, and $\ll 1\%$ of the tracers are in star-forming gas, stars, and black holes, respectively. Due to the significant variability and complexity in the evolution with time of individual tracer particles (see Appendix \ref{sec:appendix}), we focus our analysis on groups of tracers and the corresponding evolution with time of distributions of the physical quantities defined below. 

We track the position of the central galaxy throughout the subbox using the Subbox Subhalo List developed in \cite{Nelson19}, available as a supplementary data catalog for the TNG simulation suite. Positions and velocities of tracer particles are then calculated in the rest frame of their halo's central galaxy using physical coordinates. We define the specific angular momentum of a tracer particle as 
\begin{equation}
    \mathbf{j}_{\rm{tracer}} = (\mathbf{r-r}_{\rm{center}}) \times (\mathbf{v-v}_{\rm{gal}})
\end{equation}
where $\mathbf{r}_{\rm{center}}$ is the position of the center of the galaxy and $\mathbf{v}_{\rm{gal}}$ is the approximate velocity of the center of mass of the galaxy.\footnote{The Group and Subhalo catalogs that define which particles are part of which halos and galaxies, respectively, do not contain the galaxies' center-of-mass velocities and, in fact, do not exist for the subboxes at all, so we approximate the galaxy's center-of-mass velocity at every snapshot N as $(\mathbf{r_{\rm{center},N}-\mathbf{r_{\rm{center},N-1}}}) / (t_{\rm{N}} - t_{\rm{N-1}})$, where $t_{\rm{N}}$ is the age of the Universe at snapshot N.} Pressure ($P$) and entropy ($K$) are defined in terms of the temperature ($T$) and number density ($n$), which are calculated directly from the simulation, as follows:
\begin{equation}
    P = k_BTn 
\end{equation}
\begin{equation}
     K = Pn^{-5/3}
\end{equation}

\section{Results} \label{sec:results}

We first present, in Section \ref{sec:halomixingtimescales}, the typical timescale over which gas in the CGM ``mixes'' with the galaxy. Then, in Sections \ref{sec:phasetimescales} and \ref{sec:kinematictimescales}, we analyze thermodynamic and kinematic properties of the CGM over timescales shorter than this mixing timescale to understand how gas in the CGM evolves. Finally, in Section \ref{sec:iontimescales} we describe our main results in terms of ions commonly observed in quasar absorption spectra and provide evolutionary contexts for those observations. 

\subsection{CGM--Galaxy Mixing Timescales} \label{sec:halomixingtimescales}

\begin{figure}
\fig{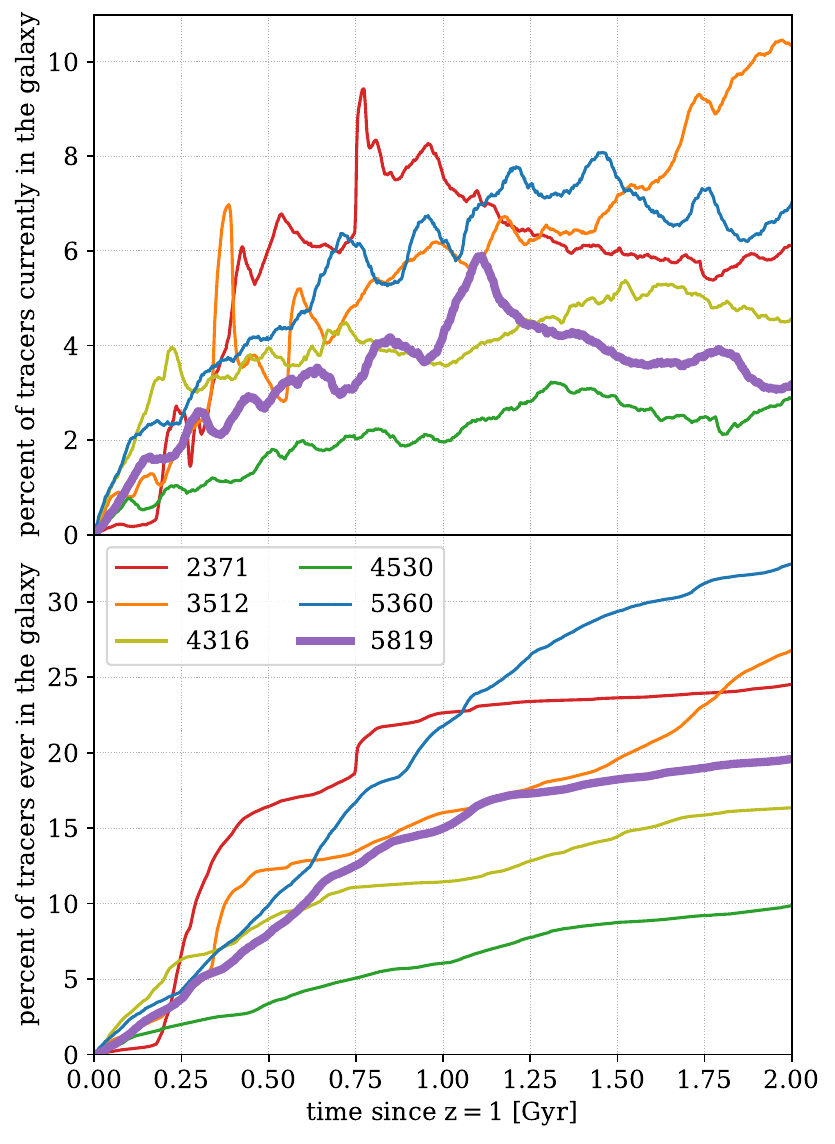}{0.48\textwidth}{}
\vspace{-25pt}
\caption{The instantaneous (top) and cumulative (bottom) percentage of the $z=1$ CGM that ever enters the galaxy as a function of time for all six halos in our sample. The thick purple line is halo5819 (see Table \ref{tab:halos}), which is used in subsequent figures.
}
\label{f:percgalaxy}
\end{figure}

We begin our investigation into the timescales of CGM evolution by establishing how quickly the CGM and ISM of the galaxy exchange mass. With the large number of snapshots we have available in the subbox, we are able to see variability in this mass exchange over very short timescales of a few Myr in addition to typical cosmological timescales of Gyr. In Figure \ref{f:percgalaxy}, we show in the top panel the percentage of the $z=1$ CGM of each of the six halos in our sample that is within its central galaxy at a later time. We note that while tracers are in the galaxy, most of them exist as either star-forming gas or stars, so we do not exclude those states from Figure \ref{f:percgalaxy}, unlike all subsequent figures. This instantaneous percentage displays clear oscillations in all six halos on periods of $\sim150$ Myr, indicative of gas that is regularly flowing nearby the galaxy. This percentage generally appears to level off after $\sim1$ Gyr, suggesting that the flow of gas between the CGM and the galaxy reaches a relatively stable equilibrium after that time. Even for the one halo in which the percentage continuously grows for all 2 Gyr, the amount of CGM gas present in the galaxy at once only ever reaches 10\%. Thus, the vast majority of the $z=1$ CGM spends most of the following 2 Gyr in the CGM. 

In the bottom panel of Figure \ref{f:percgalaxy} we show the integrated version of the top panel, thus quantifying how much of the $z=1$ CGM ever joins the galaxy in 2 Gyr. After an initial $\approx250$ Myr, this cumulative percentage at any given time is at least twice as large as the instantaneous percentage at the same time, indicating that at least half of all of the gas that accretes onto the galaxy eventually returns to the CGM. Unlike the top panel, most of the tracers that leave the galaxy and join the CGM do so in the gas phase and are \emph{not} star-forming gas or stars. Thus, in addition to retaining most of its mass, the CGM is also constantly being fed by gas reemerging from the galaxy. Furthermore, since the cumulative percentage in the bottom panel is always increasing, even for halos where the instantaneous percentage in the top panel eventually flattens, the galaxy is also being fed by CGM gas that has not accreted onto the galaxy before. Taken all together, this exchange between the galaxy and CGM reveals a system in which gas is continually processed on multi-Gyr timescales either entirely within the CGM or by briefly joining the galaxy before ending up in the CGM again. For the next three subsections, we seek to identify more specific timescales associated with this evolutionary behavior tied to different quantities, processes, and observables. 

\subsection{CGM Thermodynamic Evolutionary Timescales} \label{sec:phasetimescales}

\begin{figure*}
\fig{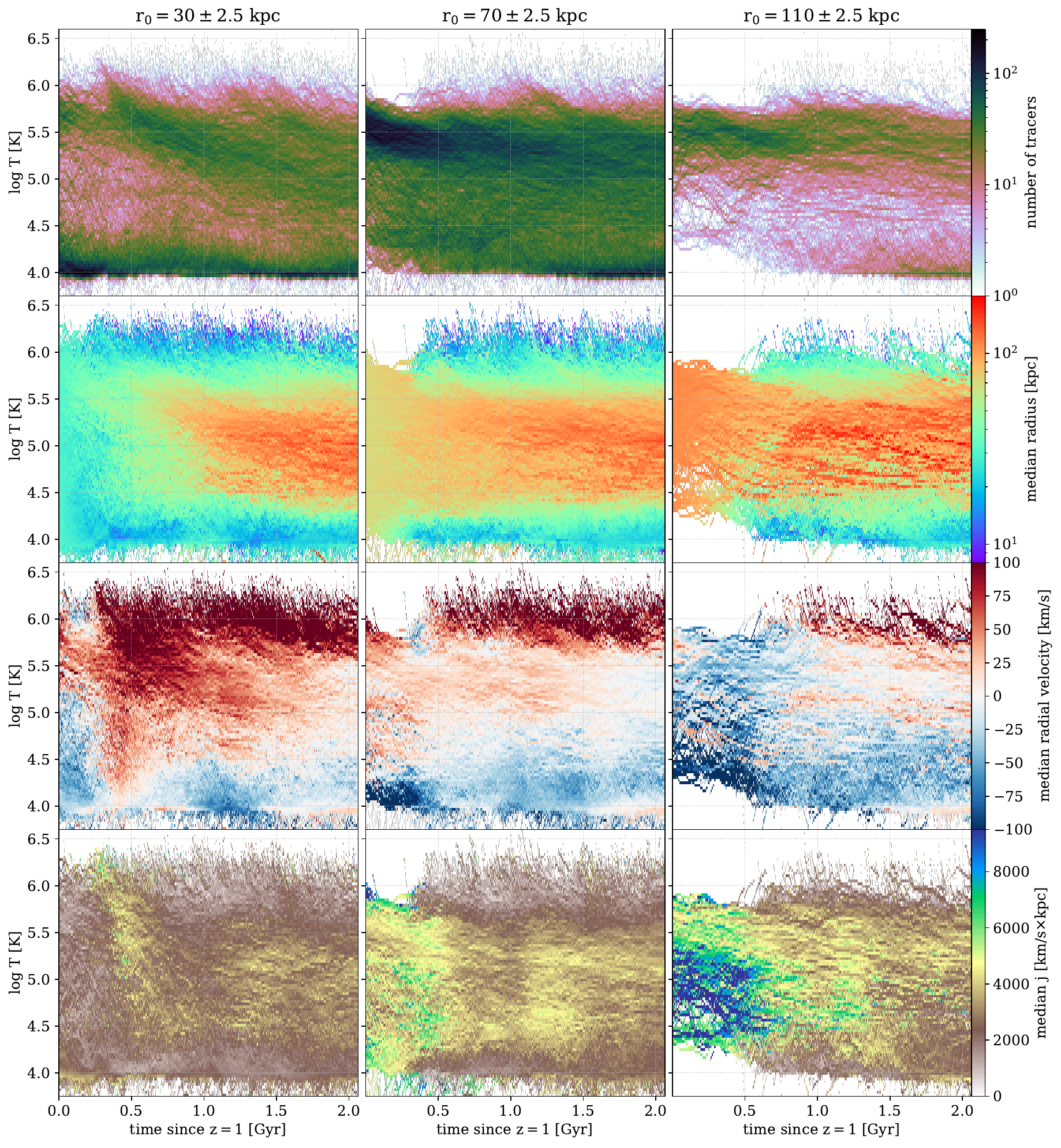}{0.99\textwidth}{}
\vspace{-15pt}
\caption{The temperature distribution of CGM gas in halo5819 with an initial radius between 27.5 and 32.5 kpc (left), between 67.5 and 72.5 kpc (middle), and between 107.5 and 112.5 kpc (right) at $z=1$ as a function of time. All distances are in physical coordinates. In the top row, the color shows the number of tracers, which is proportional to the mass, in each temperature bin at any given time. The second, third, and fourth rows are colored by the median radius, radial velocity, and specific angular momentum, respectively. By construction, all gas in each panel of the second row starts with the same color. This figure shows that CGM gas rapidly separates into a rising supernovae-heated warm--hot phase and a sinking cold phase that preferentially resides near the disk.
}
\label{f:T_dist}
\end{figure*}

\begin{figure*}
\fig{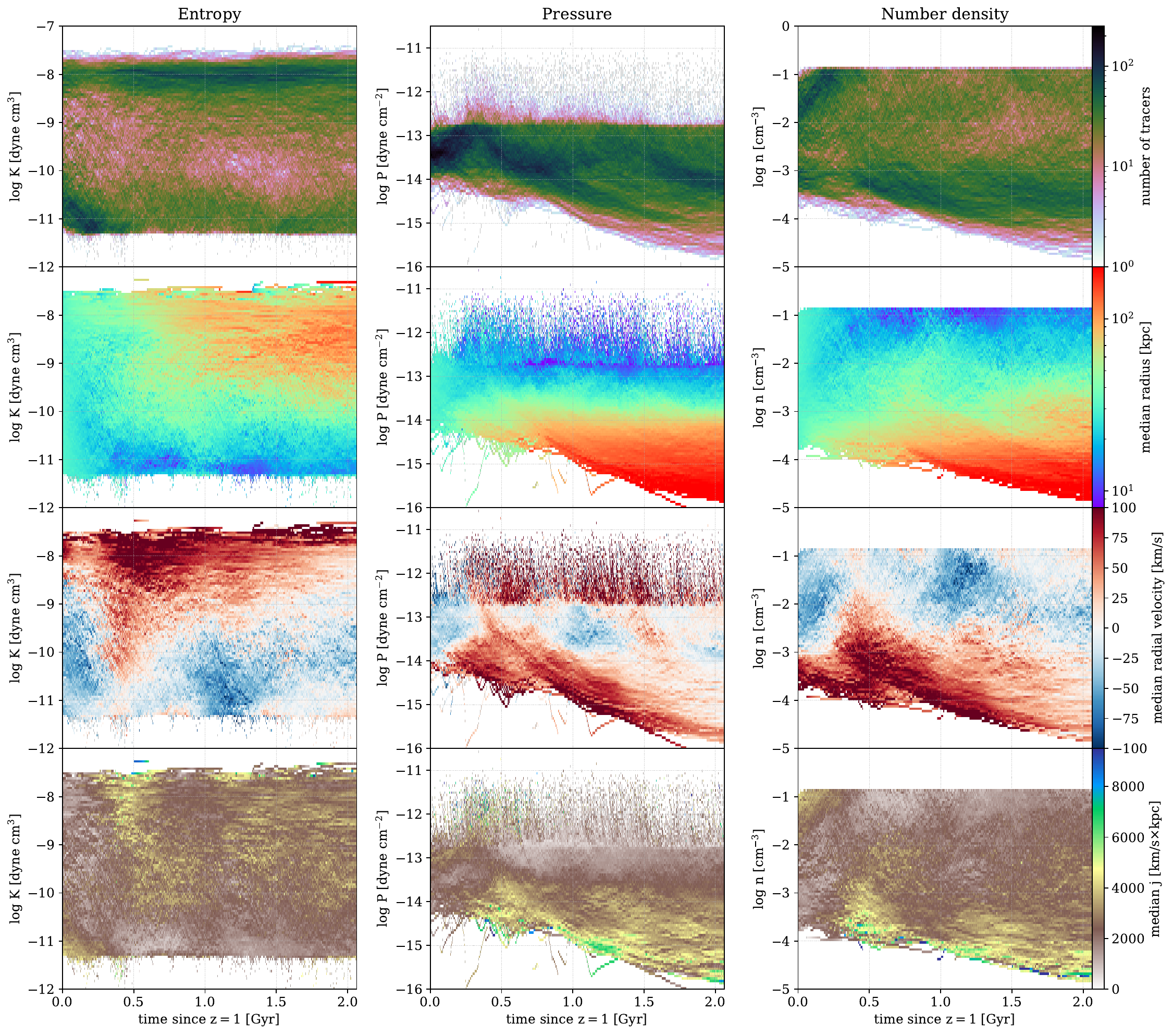}{0.99\textwidth}{}
\vspace{-15pt}
\caption{Entropy (left column), pressure (middle column), and density (right column) distributions of CGM gas from the same halo as Figure \ref{f:T_dist} at an initial selection radius between 27.5 and 32.5 kpc, colored by total mass (first row), median radius (second row), median radial velocity (third row), and median specific angular momentum (fourth row). This figure shows that two persistent entropy phases emerge, with the low (high) entropy sinking to higher (lower) densities and pressures.
}
\label{f:KPn_dist}
\end{figure*}

Having established a minimum ($\sim 2$ Gyr) length of time for which the majority of the $z=1$ CGM evolves outside of the galaxy, we now examine detailed aspects of that evolution by plotting distributions of thermodynamic quantities over time. We start in Figure \ref{f:T_dist} with temperature ($T$) using halo5819 from Table \ref{tab:halos}, which has a CGM that is relatively undisturbed by mergers and is therefore useful for establishing a base case. We separate CGM gas tracers into three regimes based on their initial radii at $z=1$ and follow those sets of tracers over time. At an initial radius $r_0 \approx 30$ kpc (left panel), the CGM is already separated into two temperature phases: a cold phase around the temperature floor of $10^4 \, \rm{K}$, and a warm--hot phase around $10^{5.5} \, \rm{K}$, approximately the halo's virial temperature. Over time, we see that the cold phase maintains a relatively constant number of tracers (a proxy for mass) and temperature, while the warm--hot phase appears to slowly and coherently cool to temperatures around $10^5 \, \rm{K}$ and periodically get replenished with gas as hot as $10^6 \, \rm{K}$. This warm--hot phase has a much wider temperature distribution that also changes significantly with time, unlike the cold phase. At larger initial radii (middle and right panels), this general behavior changes in two key ways. First, the majority of the gas begins in the warm--hot phase rather than split roughly equally between warm--hot and cold, and the warm--hot phase remains the dominant phase in terms of mass over the entire $2$ Gyr time range. Second, gas takes as long as $1$ Gyr to enter the cold phase. Together these radial differences suggest that the warm--hot and cold phases themselves have different radial distributions. 

We see this very clearly in the second row of Figure \ref{f:T_dist}, which shows the same temperature distributions as a function of time, but now color-coded by the median radius rather than the number of tracer particles. Examining these panels in conjunction with the top panels of Figure \ref{f:T_dist} reveals many key aspects of the formation and evolution of these two temperature phases. In all three radial regimes, the two phases end up being separated by radius, with the cold phase forming and largely settling between $10$ and $30$ kpc and the warm--hot phase moving outward from its initial radius to distances $>100$ kpc from the central galaxy, where it remains. These radial distributions as a function of time also help explain evolutionary differences between the initial radial selections. For example, while a large fraction of the $r_0\approx30$ kpc selection starts out in the cold phase at $\approx10^4 \, \rm{K}$, it takes some amount of time for any gas from the other two radial selections to enter the cold phase: for gas starting at $r_0\approx70$ kpc, it takes $\approx 200$ Myr for any of the gas to cool all the way to $10^4 \, \rm{K}$, and for gas starting at $r_0\approx110$ kpc, it takes even longer ($\approx 600$ Myr). Evidently, this is because such gas preferentially forms and remains close to the galaxy, so gas selected farther out in the halo must travel for a longer time to the inner CGM in order join the cold phase. This description is further supported by the third row of panels in Figure \ref{f:T_dist}, which show the median radial velocity: hotter gas above $10^5 \, \rm{K}$ initially moves both outward (for smaller initial radii) and inward (for large initial radii), while cold gas near $10^4 \, \rm{K}$ initially moves only inward. After a few hundred Myr, even though all the gas tracers are still moving radially, the radial velocity distributions reach a kind of steady state and essentially become time independent with little further overall radial evolution. We consider observational implications of this two-phase dichotomy in Section \ref{sec:iontimescales}.

In addition to the two main warm--hot and cold phases, there is also a subset of very hot ($T > 10^6 \, \rm{K}$) gas that always develops concurrently and cospatially with the cold phase. This gas, though containing very little mass, has very high radial velocities ($\gtrsim100 \, \rm{km} \, \rm{s^{-1}}$) as shown in the third row of Figure \ref{f:T_dist}, consistent with being launched by galactic winds from energy injected by supernovae. In examining individual tracers, we find that some of this material cools quickly, loses radial velocity, and reenters the cold phase as it falls back toward the galaxy, while the rest remains in the warm--hot phase as it continues expanding outward and cools down more slowly. The consistent appearance of this phase of gas, along with the cold phase at small radii, regardless of the initial phase and radius of that gas, demonstrates that some fraction of halo gas is always cycling through the CGM and getting close enough to the galaxy to be affected by feedback, as suggested by Figure \ref{f:percgalaxy}, thus helping to fuel mixing between hot and cold gas. We speculate that the cospatial nature of the very hot gas and the cold phase indicates that there could regularly be mixing and condensation of small clouds occurring, particularly in the inner regions of outflows, that may require a higher spatial resolution than found in these simulations to fully resolve. We also note here that there have been detections of a supervirial hot phase in the Milky Way's CGM with temperatures above $10^6 \, \rm{K}$ \citep{Das19,Das21,McClain24} and that this feedback-heated gas close to the galaxy may be a possible origin of this detected phase, though examining this connection in detail is beyond the scope of this text.

In the fourth row of panels in Figure \ref{f:T_dist} we show the median specific angular momentum ($j$) of this CGM gas with respect to the galaxy. Gas that starts at or moves to larger radii generally has correspondingly larger $j$. Gas that has large negative radial velocities tends to also lose angular momentum as it moves toward the galaxy. Over time, the warm--hot phase has higher $j$ than both the cold phase, due to the cold phase's proximity to the galaxy, and the very hot wind, due to the wind's primarily radial velocities. This means that the warm--hot phase, which is sourced from gas originating at radii throughout the CGM, must have significant nonradial velocities while the fast-moving wind joins it and slowly cooling and inflowing gas leaves it. This has been examined by works like \cite{Sormani18} and \cite{Sobacchi19}, which present models of rotating hot halos, and it is also supported by \cite{DeFelippis20}, who find that hotter gas in the CGM often rotates coherently like the cold gas, even at large radii, as well as \cite{Stern24}, who demonstrate that rotating hot CGM gas can accrete toward the galaxy before cooling at the disk--halo interface.
 
\begin{figure*}
\fig{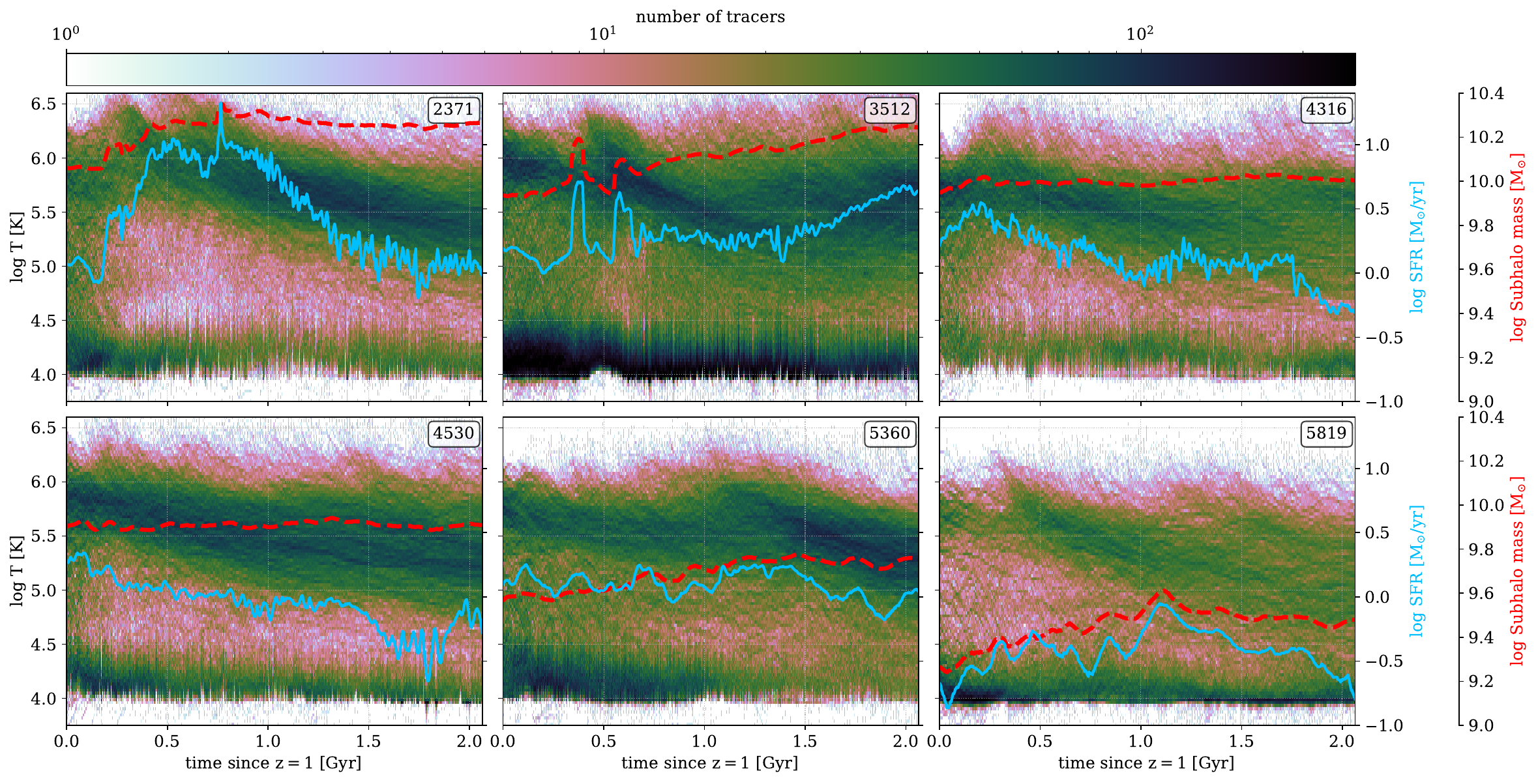}{0.99\textwidth}{}
\vspace{-15pt}
\caption{Temperature distributions of CGM gas at an initial radius of $30\pm 2.5$ kpc as defined in Figure \ref{f:T_dist} for all six halos in our sample. The central galaxy's SFR and total baryonic mass as functions of time are overplotted on their corresponding halo's panel with solid blue and dashed red lines, respectively. Each panel shows the corresponding halo ID from Table \ref{tab:halos}. We see in this figure that the two-phase temperature structure of the CGM appears in all halos, with differences driven in part by variations in star formation history and mergers.
}
\label{f:galaxy_properties}
\end{figure*}

As we have shown, the evolutionary behavior of the gas temperature is very similar at all radii; we now focus on the evolution of other thermal quantities possessed by gas at an initial radius of $\approx30$ kpc, noting that there is little variation with initial radius. In Figure \ref{f:KPn_dist}, we plot the entropy ($K$; left column), pressure ($P$; middle column), and number density ($n$; right column) of this gas in the same way as in Figure \ref{f:T_dist}. In the entropy plots, the two main temperature phases show up very clearly as a high-entropy phase and a low-entropy phase, which correspond to the warm--hot and cold phases, respectively. However, unlike the temperature, the median entropy of both phases is constant over the entire 2 Gyr, with very little mass occupying any entropy values between them at any given time. This indicates that the cooling processes that drive the evolution of the warm--hot phase are adiabatic (i.e. at constant entropy) rather than radiative \citep[see the isentropic model studied in][]{Faerman20}. The two entropy phases are driven apart over time, with the higher-entropy phase rising to larger radii in the halo and the lower-entropy phase falling to lower radii. At the same initial radius, the evolutions of both radial velocity and angular momentum within the high- and low-entropy phases are very similar to their corresponding evolutions within the warm--hot and cold temperature phases, respectively. 

The pressure and density distributions strongly resemble each other and, unlike temperature and entropy, do not have or develop any clear bimodalities. The very hot, high-velocity gas from Figure \ref{f:T_dist} is seen as persistent high-pressure spikes. Apparent truncations in density (above $\approx 0.13$ cm$^{-3}$) and pressure (above $\approx10^{-12.6}$ dynes cm$^{-2}$) correspond to the values assigned to those quantities for star-forming gas and within the galaxy--CGM interface.\footnote{The TNG physics model uses an effective equation of state to define the pressure and temperature of star-forming gas, meaning that their numerical values are unphysical \citep[see][]{Springel03}. As is the case for stars and black holes, we exclude tracers currently in the star-forming gas phase until they become ``regular'' gas again.} Over time, both distributions widen, with gas reaching lower and lower pressures and densities, which is also where the angular momentum and radial velocities are the highest. Furthermore, neither pressure nor density distributions at a given radius change substantially with time (as shown by the mostly horizontal color bands in the second row of Figure \ref{f:KPn_dist}), indicating that there is a relatively stable profile throughout the halo that a subset of gas starting at $\approx30$ kpc (or other larger radii) naturally follows as it segregates itself into the two previously identified temperature and entropy phases. We note here that \cite{Dutta22} studied a particular mechanism by which gas can cool out of a hotter CGM with a steady-state pressure gradient. The relatively steady pressure gradient, along with the negative radial velocities at smaller radii that we see in Figure \ref{f:KPn_dist}, may be a tentative signature of a process similar to this.

The trends in thermodynamic properties we have shown so far have been for one halo, but these results are qualitatively true of the other halos in our sample. For example, in Figure \ref{f:galaxy_properties} we show the evolution of the temperature distribution for the gas with an initial radius of $\approx30$ kpc for all six halos in our sample (the bottom right panel is the same as the top left panel of Figure \ref{f:T_dist}). In all cases, a two-phase temperature distribution exists or develops in $<200$ Myr, and we find the same behavior for the entropy as well. The only significant difference  in the gas distributions between halos is the relative amount of mass in the two phases, suggesting possible effects from the different evolutionary histories and cosmological environments of the halos. 

To test one aspect of this, we show in Figure \ref{f:galaxy_properties} each central galaxy's instantaneous SFR and total baryonic mass as a function of time. It appears, especially based on the two halos in the middle column, that having a higher SFR for a longer period of time results in more mass in the CGM at all temperatures, including at intermediate temperatures where gas is transitioning between the two main phases. Furthermore, short periods of high SFR are associated with more gas joining the warm--hot phase. Some of this is certainly due to increased stellar feedback from the galaxies when the SFR is higher, although it is also clear that mergers with satellite galaxies, which show up as spikes on the galaxy's mass over time (particularly in the top left and middle panels), may drive this evolution as much as steady stellar feedback does by providing a large amount of gas at once for the galaxy and CGM to begin processing. Still, satellite mergers are rare, and the evolutionary behavior we have identified, including the presence of a hot high-velocity wind, occurs even when the SFR is consistently below $1 \, M_{\odot} \, \rm{yr^{-1}}$, as in the bottom right panel, again highlighting the ubiquity of that behavior. We also note here that many of the halos display variations in their SFR that are comparable to the variations in the amount of tracers in the central galaxy shown in Figure \ref{f:percgalaxy}, which suggests a possible connection between waves of gas accretion and subsequent star formation, though we defer further analysis of this variation to future work. 

\subsection{CGM Kinematic Evolutionary Timescales} \label{sec:kinematictimescales}

\begin{figure*}
\fig{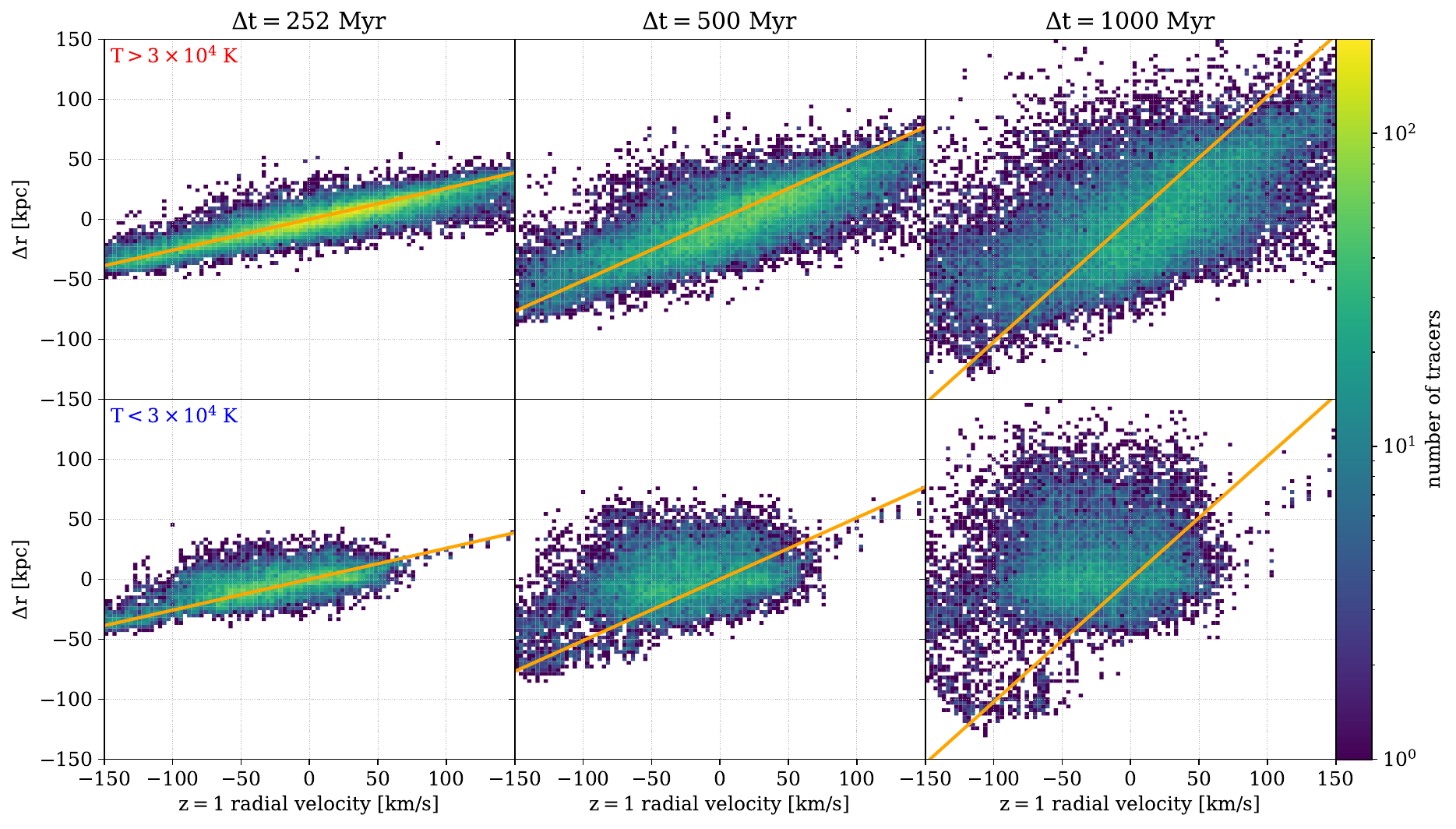}{0.99\textwidth}{}
\vspace{-15pt}
\caption{The change in radius of tracers in the $z=1$ CGM of halo5819 as a function of their initial radial velocities. Tracers are separated by whether their initial temperatures are above (top row) or below (bottom row) $3\times10^4 \, \rm{K}$. Each column shows a different amount of time after $z=1$. By design, the tracers cannot move horizontally across panels. The orange lines show the change in radius after the associated elapsed time assuming a constant radial velocity ($\Delta r = v_r(z=1) \times \Delta t$), and the slopes of these lines are $\Delta t$. This figure shows that CGM gas initially evolves approximately ballistically for $\sim250$ Myr before feedback and mixing erase the memory of its initial radial velocity.
}
\label{f:deltar_vr}
\end{figure*}

We now seek to understand whether the evolutionary behavior of thermal quantities in the CGM is accompanied by evolution of kinematic quantities. We have already identified radial velocity as a key difference separating the warm--hot and cold temperature phases that emerge over time, so it is natural to examine in more detail whether and how the radial velocity of CGM gas is influencing its evolution. We start in Figure \ref{f:deltar_vr} by plotting the change in radius of CGM tracers in a single halo over multiple time periods as a function of their initial (namely, $z=1$) radial velocities. We also plot a straight line on top of the tracers showing the change in radius assuming that the initial radial velocity remains unchanged. In the leftmost panels, where $\approx250$ Myr have elapsed since $z=1$, the tracers largely follow this line, whether they originated in the cold phase as seen in Figure \ref{f:T_dist} with a temperature $\leq3\times10^4 \, \rm{K}$, or the warm--hot phase with a temperature $\geq 3\times10^4 \, \rm{K}$. Gas in the cold phase has slightly more scatter and is biased toward lower radial velocities in general, but both phases appear similarly centered on the line, indicating that most of the mass in the CGM has moved at its initial radial velocity for $250$ Myr. After a further $250$ Myr, as shown in the middle panels, the scatter of the warm--hot phase around the constant radial velocity line has increased, but the line is still predicting the mean change in radius fairly well. Most of the cold phase, on the other hand, has not moved inward or outward from its position after the first $250$ Myr, and the distribution of that gas has seemingly decoupled from the constant $v_r$ line. After $1$ Gyr, both phases have evolved such that a constant radial velocity is not predictive of the tracers' locations, although the distribution of the warm--hot phase still retains an upward slope, indicating that the average radial evolution of this phase is relatively slow. The behavior of the cold phase implies that there is a characteristic timescale between $\approx250$ and $500$ Myr after which gas that starts out in the cold phase ``forgets'' its radial velocity history completely as it is reprocessed via feedback and accretion. We highlight here the works of \cite{Pandya23} and \cite{Voit24}, which examine kinematic and energetic relationships between different phases of gas in the CGM as a function of time and identify multiple processes that can affect those relationships, including feedback from supernovae, mass loading, and turbulence. As the methodologies of these works are very different from ours and each other, we will perform a detailed comparison in future work.

\begin{figure}
\fig{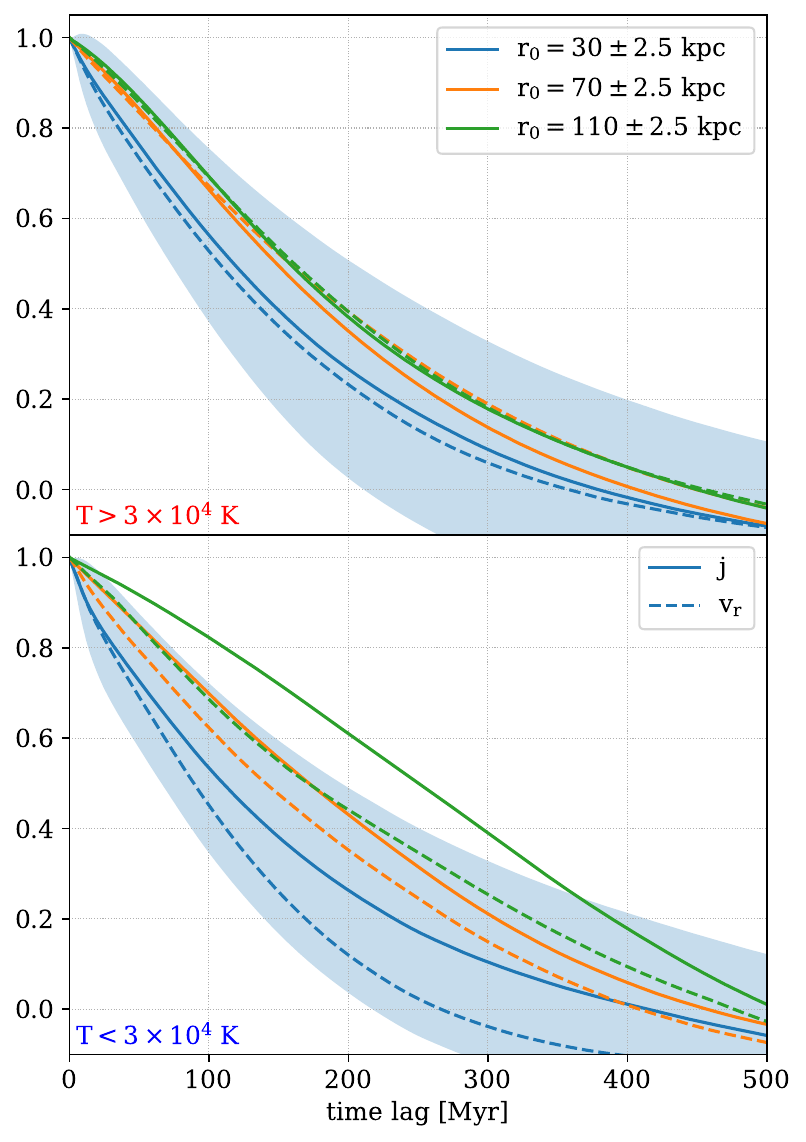}{0.48\textwidth}{}
\vspace{-25pt}
\caption{The mean autocorrelations of the radial velocities and specific angular momenta of $z=1$ CGM tracers from halo5819 starting at three different initial radii. Tracers are separated by whether their initial temperatures are above (top panel) or below (bottom panel) $3\times10^4 \,\rm{K}$. The $1\sigma$ spread for $j$ at $r_0=30$ kpc is shown in both panels and is comparable to all other radii and to that of $v_r$.
}
\label{f:autocorrelation}
\end{figure}

In order to quantify this kinematic timescale, we compute the autocorrelations of two key kinematic quantities of CGM gas: the radial velocity ($v_r$) and specific angular momentum ($j$) with respect to the galaxy. In doing so, we can identify both whether and for how long $v_r$ and $j$ depend on their past values. We show these autocorrelations in Figure \ref{f:autocorrelation}, separated into the same initially warm--hot and cold phases shown in Figure \ref{f:deltar_vr}. Gas beginning in the warm--hot phase has essentially the same autocorrelation profile for both $v_r$ and $j$, and there is also little variation with initial radius: as the time lag increases from zero, the autocorrelation declines from its initial value of $1$ and reaches a value of $0$ at around $400$ Myr. We interpret this as the time at which the initial values of both $v_r$ and $j$ can no longer predict their ``current'' values. The lack of significant variation with radius is consistent with the ubiquity of the warm--hot phase of the CGM as seen in Section \ref{sec:phasetimescales}. 

Gas starting in the cold phase has qualitatively similar $v_r$ and $j$ autocorrelations that peak at zero time lag and decline to a value of $0$ after $\sim400$ Myr, but there are distinctions between the two kinematic quantities, as well as variation with initial radius. At any initial radius, the autocorrelation of $v_r$ is strictly less than the autocorrelation of $j$, and for both $v_r$ and $j$ the autocorrelation is smaller at lower initial radii. We can understand this by noting that cold gas is more likely to behave ballistically than hot gas, where, in absence of outside forces, $j$ is a conserved quantity while $v_r$ is not.\footnote{This is an admittedly simplified picture, as it ignores processes that can affect cold gas formation and survival like condensation and mixing that likely contribute to the large spread in the autocorrelations.} Thus, we would expect $v_r$ to decorrelate faster than $j$ at all radii. The trend with radius of both quantities is likely due to the proximity of the feedback from the galaxy, which will ``erase'' initial kinematic quantities associated with the cold gas more effectively by heating and pushing it out when it is closer to the galaxy to begin with, resulting in more significant mixing and turbulence, whereas gas farther out is in a generally calmer environment that evolves more slowly and is less affected by those processes. It is also possible that the cold phase forms in multiple ways (e.g. precipitation from the warm--hot phase, or turbulent mixing) while feedback or accretion in the inner CGM is occurring. Studying these processes in greater detail through idealized cloud crushing simulations such as \cite{Gronke18}, or recent efforts like \cite{Ramesh26} that include cosmological environments, reveals that resolving much smaller scale structure is necessary to understanding in detail how feedback can alter the kinematic properties of gas as it is thrown out of the galaxy. As we are limited by the cosmological resolution of TNG, we defer further analysis of this evolution to future work. 

\subsection{CGM ``Observed'' Phase Evolutionary Timescales} \label{sec:iontimescales}

In this subsection, we attempt to translate the CGM evolutionary behavior we have described so far using physical quantities to quantities commonly measured in observations of the CGM using quasar absorption lines. We do this with the goal of providing evolutionary context to $z\approx1$ observations from surveys like MEGAFLOW \citep{Bouche25} by characterizing how the material in such systems is likely to change, if at all, and where it is likely to end up. We focus on three temperature and density regimes associated with three different ions: \MgII, which traces cold dense gas around $T=10^4 \, \rm{K}$, \CIV, which traces warmer gas around $T=10^5 \, \rm{K}$, and \OVI, which traces hotter and more diffuse gas at $T=10^{5.5} \, \rm{K}$. We use Figure 6 from \cite{Tumlinson17} as a guide for defining the temperature and number density ranges associated with the absorption of these three ions, and we refer to gas inside these ranges as \MgII--like ($T < 3\times10^4 \, \rm{K}, \; 10^{-2} \, \rm{cm^{-3}} < n < 0.13 \, \rm{cm}^{-3}$), \CIV--like ($10^{4.75} \, \rm{K} < T < 10^{5.25} \, \rm{K}, \; n < 10^{-2} \, \rm{cm^{-3}}$), and \OVI--like ($10^{5.25} \, \rm{K} < T < 10^{5.75} \, \rm{K}, \; n < 10^{-3} \, \rm{cm^{-3}}$).

\begin{figure*}
\fig{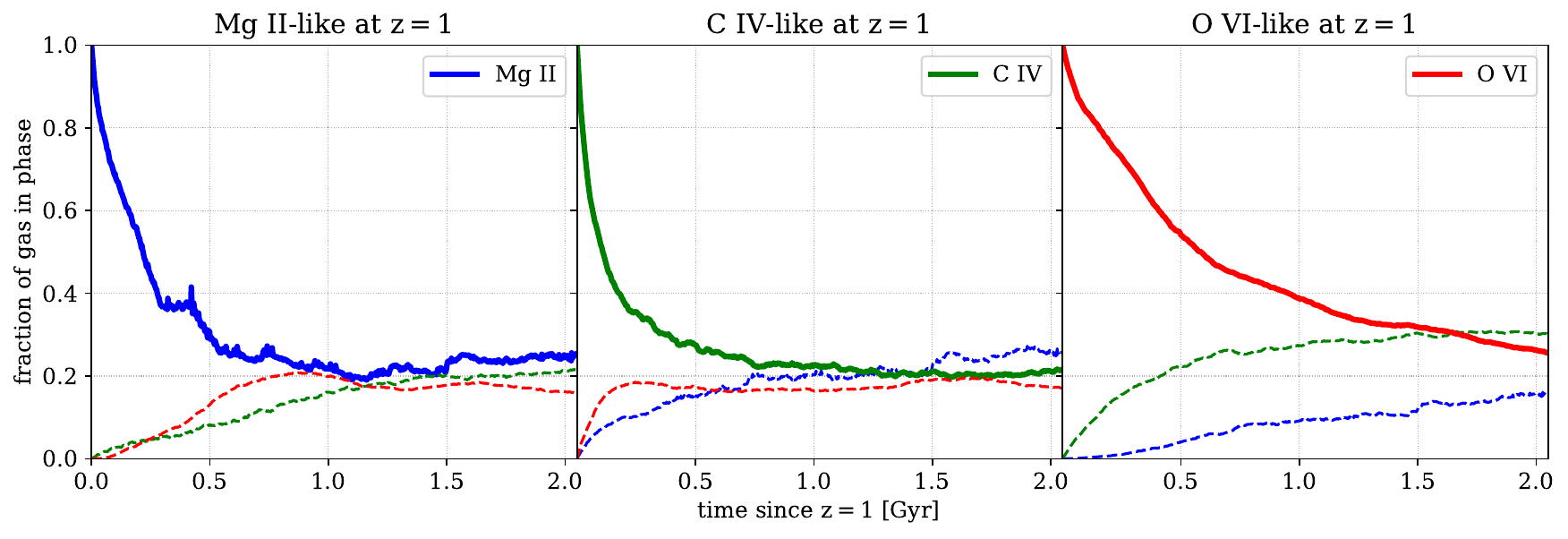}{\textwidth}{}
\vspace{-25pt}
\caption{The fraction of all $z=1$ CGM gas from halo5819 associated with three different ions as a function of time, separated by being initially associated with \MgII{} (left), \CIV{} (middle), and \OVI{} (right). By construction, these fractions sum to $1$ at the initial time because the gas is initially selected to be associated with one ion, but at later times they do not necessarily sum to $1$ because some gas evolves outside the temperature and density ranges associated with all three ions. Gas that is initially \MgII--like or \CIV--like mixes with other phases at similar rates, but gas that is initially \OVI--like mixes with other phases significantly more slowly.
}
\label{f:gasphasefrac}
\end{figure*}

First, in Figure \ref{f:gasphasefrac}, we attempt to understand how quickly gas transitions between these three observed phases. To do this, we show the fraction of all CGM gas from halo5819 initially associated with a particular ion at $z=1$ that remains associated with that ion, as well as the fractions that become associated with the other two ions, as a function of time. A small amount of the gas that starts in any ion changes its properties enough to be observed as a different ion almost immediately (i.e. $<100$ Myr). However, the precise behavior of this transition over time is very ion dependent. \MgII--like gas transitions to both \CIV--like and \OVI--like gas in roughly equal proportion over time, and after $1$ Gyr it is roughly equally distributed among all three types of gas. \CIV--like gas behaves very similarly by distributing itself equally among all three types and reaching a completely mixed state after $\approx1$ Gyr. On the other hand, the \OVI--like gas appears more stable: after $1$ Gyr it is still the most common type of gas. There is also a clear path of transition first to \CIV--like gas and then to \MgII--like gas as evidenced by the initial rise of the \CIV--like fraction in the first $\approx200$ Myr compared to the delayed rise of the \MgII--like fraction in the same time period. In other words, it takes longer for \OVI--like gas to transition to \MgII--like gas than it does for \MgII--like gas to transition to \OVI--like gas. 

\begin{figure}
\fig{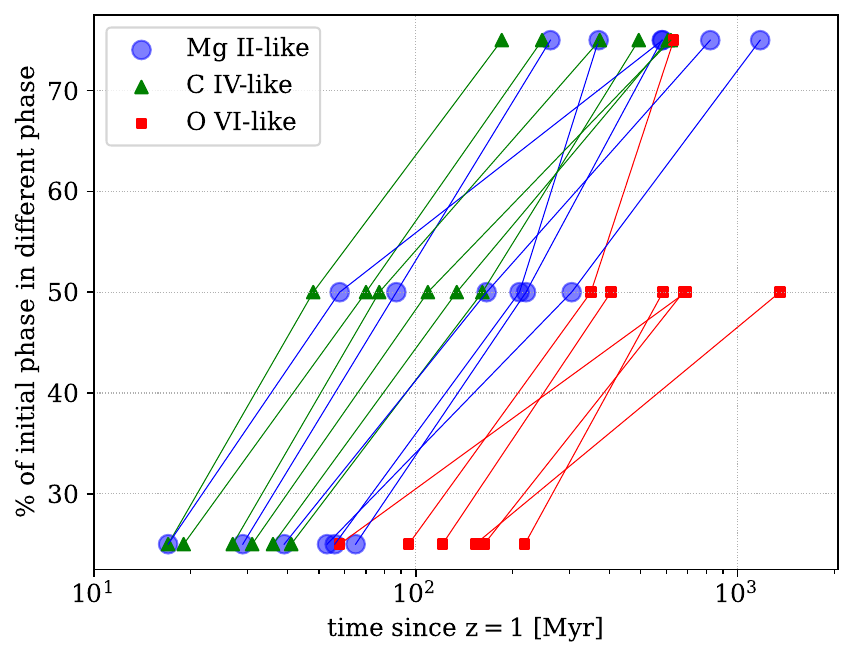}{0.48\textwidth}{}
\vspace{-25pt}
\caption{A complement to Figure \ref{f:gasphasefrac} showing different timescales to reach multiple levels of ion phase mixing for all six halos in our sample.
}
\label{f:gasphasestability}
\end{figure}

We generalize this from a single halo to our entire sample of six halos in Figure \ref{f:gasphasestability}, which shows the time elapsed for one-quarter, one-half, and three-quarters of the different ion-like gases at $z=1$ to no longer be associated with that ion. As expected from Figure \ref{f:gasphasefrac}, both \MgII--like and \CIV--like gas transitions out of its initial phase faster than \OVI--like gas by about a factor of three. The variation from halo to halo of that transition timescale for a given ion is also about a factor of three, indicating that while the basic phase evolution is similar across halos, the detailed transition rates depend on the specific properties of each system. Differences in star formation histories, merger activity, and the surrounding cosmological environment modify the heating, cooling, and radial transport and mixing of CGM gas. While half of the \MgII--like and \CIV--like gas is in a different phase after $100-200$ Myr, it takes at least $300$ Myr for a comparable amount of \OVI--like gas to transition to a different phase, similar to the kinematic timescales described in Section \ref{sec:kinematictimescales}. The relative stability of the \OVI--like phase suggests that it could be directly related to the warm--hot phase that CGM gas tends to evolve into identified in Section \ref{sec:phasetimescales}. 

\begin{figure*}
\fig{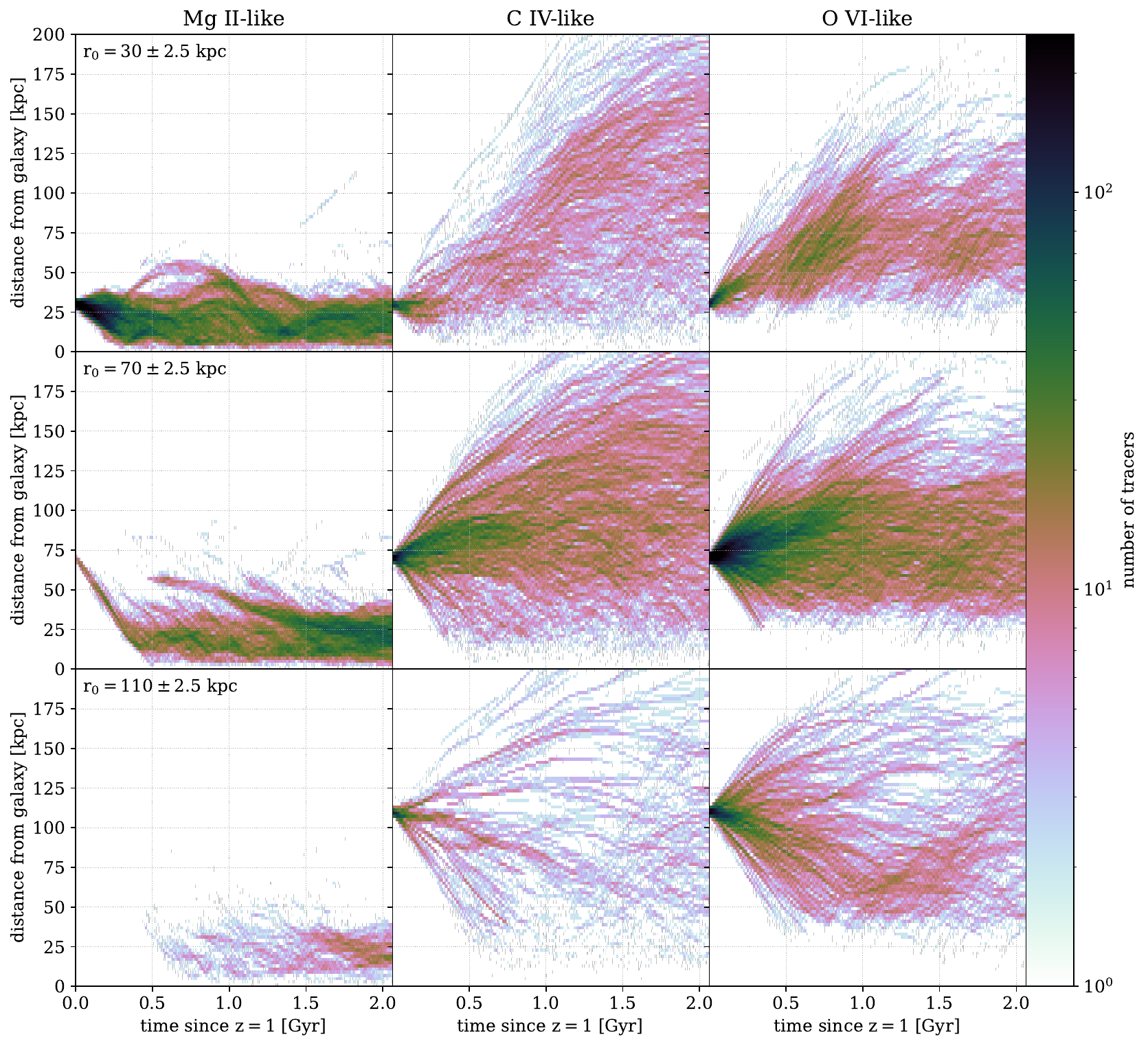}{0.99\textwidth}{}
\vspace{-15pt}
\caption{The location of CGM gas from halo5819 with an initial radius of $\approx30$ kpc (top), $\approx70$ kpc (middle), and $\approx110$ kpc (bottom) at $z=1$ as a function of time separated by the observed phase of that gas. The coldest and densest gas (\MgII--like; left panel) is always seen closer to the galaxy, while warmer and more diffuse phases (\CIV--like and \OVI--like; middle and right panels, respectively) appear as gas moves outward from its initial radius, both toward and away from the galaxy.
}
\label{f:ion_rdist}
\end{figure*}

We further examine this idea by plotting the radial distributions of the three ion phases as a function of time in Figure \ref{f:ion_rdist} for the same three initial radii as in previous sections. We emphasize that individual tracer particles are not confined to a single ion phase and can transition between the \MgII--like, \CIV--like, and \OVI--like regimes as their temperature and density evolve. We find that each of the three ions is ``observed'' at different quantities and locations depending on where the gas originated in the CGM. \MgII--like gas always forms and exists only at small radii ($<50$ kpc) in the CGM, unless, as in the second row of Figure \ref{f:ion_rdist}, it is part of an accreting satellite or condenses out of a different phase while inflowing. \MgII--like gas that originates at a small radius, as in the top row, exhibits oscillatory behavior with time, indicating that much of this phase may be undergoing rotation near the galaxy with a period approaching $1$ Gyr (though other processes near the galaxy like mixing, turbulence, and/or bursts of supernova activity can also cause similar oscillatory behavior). We also see that the more distant the origin of the gas, the longer it takes for a \MgII--like phase to develop out of that gas, as some quantity of gas needs to move inward toward the galaxy and cool. In these ways, the \MgII--like gas phase develops with time very similarly to how the cold phase develops with time in Figure \ref{f:T_dist}.

While initial gas selections at different radii contain a mix of all the ions (except $r_0 = 110$ kpc), the \OVI--like and \MgII--like phases quickly separate and become spatially distinct. Gas with a small initial radius that is seen as \OVI--like gas is almost exclusively moving outward over time, but no matter the origin, the \OVI--like phase spreads out and is found at all radii $\gtrsim25$ kpc, essentially filling the outer CGM. Wide, $>1$ Gyr ``arcs'' can be seen in all of the \OVI--like panels, indicating that this gas tends to be slowly flowing through the CGM rather than being ejected out beyond the virial radius at high velocities, broadly consistent with the behavior of the high angular momentum warm--hot phase that slowly cools over time.

The \CIV--like phase, an intermediate phase between \MgII--like and \OVI--like, therefore represents much of the gas that passes between the two main temperature phases: it is found at lower radii than \OVI--like gas but not as low as \MgII--like gas, and it broadly evolves the same way the \OVI--like phase does, by spreading throughout the halo. However, a key difference is in the behavior at small initial radius (the first row of Figure \ref{f:ion_rdist}): while the \OVI--like gas has largely positive radial velocities (i.e., the aforementioned ``arcs'') that are pointing away from where the \MgII--like phase is, \CIV--like gas has no clear preferred radial velocity direction and more substantially overlaps the radii where the \MgII--like phase is ($\lesssim20$ kpc). In other words, it appears just as likely for this \CIV--like gas to be associated with the \OVI--like phase that is moving radially outward or the \MgII--like phase that is remaining in the inner CGM.

\section{Discussion} \label{sec:discussion}

In this section, we first examine underlying profiles of certain CGM quantities to attempt to explain the drivers of the evolution seen throughout Section \ref{sec:results}. Then, we compare our results to other recent analyses of CGM gas in simulations. Finally, we connect our results to recent observations of the CGM at $z\approx1$. 

\subsection{Physical Drivers of CGM Gas} \label{sec:drivers}

\begin{figure}
\fig{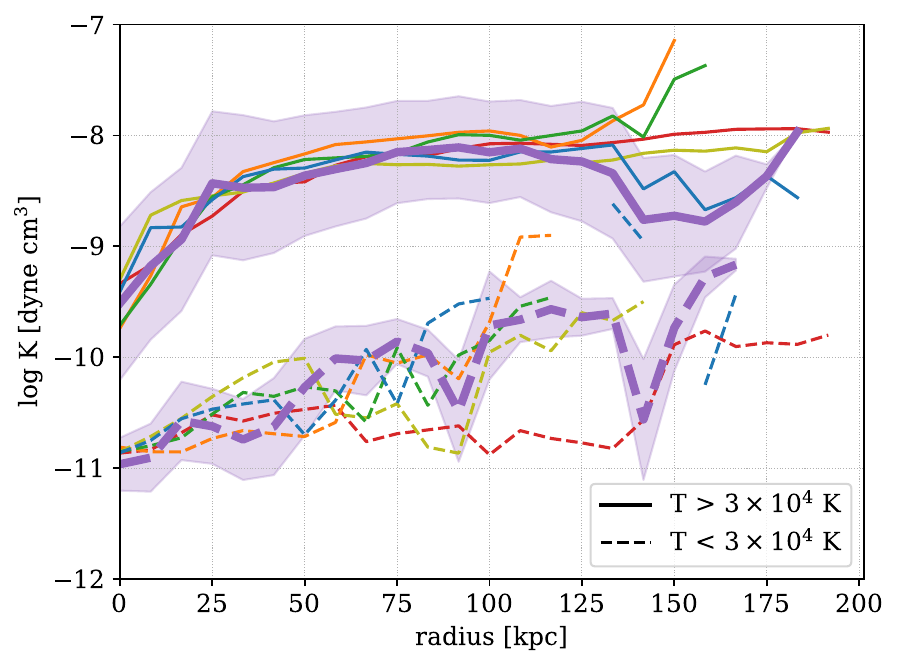}{0.48\textwidth}{}
\vspace{-25pt}
\caption{Median entropy radial profiles of the warm--hot (solid lines) and cold (dashed lines) phases of the CGM for our sample of six halos at $z=1$. The $1\sigma$ spread for halo5819 is shown and is comparable to the other halos. The warm--hot phase has a relatively flat entropy profile in the CGM of all halos beyond $\approx40$ kpc, while the cold phase's profile varies by $\sim1$ dex in the same radial range.
}
\label{f:Kprofile}
\end{figure}

We saw in Section \ref{sec:phasetimescales} that a key behavior of CGM gas that occurs regardless of its origin is the outflow of a slowly cooling phase of gas that is heated up by feedback near the galaxy. This slow temperature decrease occurs at nearly constant entropy and involves high-entropy gas adiabatically rising, a process reminiscent of convection or of an energy-driven wind like that of \cite{Chevalier85}. In the left panel of the third row in Figure~\ref{f:KPn_dist}, we see clearly that high-entropy fluid elements tend to be rising, while low-entropy fluid elements sink, consistent with this approximate analogy. We examine this in more detail in Figure \ref{f:Kprofile}, where we plot the median radial profiles of entropy for the warm--hot and cold phases referred to throughout Section \ref{sec:results} for each of the six halos in our sample at $z=1$. We see a striking difference between the two phases. All halos have essentially the same entropy profile for the warm--hot phase, which rises near the galaxy and plateaus around $40$ kpc out to the virial radius, with very little scatter. The cold phase profiles, however, vary much more significantly from halo to halo, indicative of the different number of satellite galaxies and overall amount of dense structures present in the halos' CGM. While we do not study the central galaxies' satellites in detail, it is likely that they interact somewhat with the CGM: studies such as \cite{Ghosh24} and \cite{Sparre24} show that satellites can be at least partially stripped and contribute a nonnegligible amount of cool gas as they pass through their host halo. It is also possible that the large variation in individual cold phase profiles, especially $>50$ kpc, is due to the limited resolution in the CGM, and we may expect the entropy profiles to be smoother at higher resolution \citep[see, e.g.,][]{Schneider24}. The general radial behavior of the entropy in these halos is similar to the model described in \cite{Keller20}, where buoyantly driven galactic outflows result in an increasing overall entropy profile as a function of radius. In this model, outflows tend to have smaller velocities ($<100 \, \rm{km \, s^{-1}}$) and occur over longer times (hundreds of Myr), which is qualitatively consistent with the behavior of gas with positive $v_r$ shown in Figures \ref{f:T_dist} and \ref{f:KPn_dist}. 

In our limited sample, focused on halos up to $10^{12} \, M_{\odot}$, the total black hole mass is in the range of $\approx(3-50)\times10^6 \, M_{\odot}$, which is below the mass of the supermassive black holes ($\approx10^{8} \, M_{\odot}$) that power the strong kinetic AGN feedback in the IllustrisTNG model \citep{Terrazas20}. Thus, the feedback that affects the structure and dynamics of the CGM of these galaxies is dominated by star formation and supernovae. Stellar feedback has been shown in simulations to generate fountain flows whereby material is ejected from the galaxy into the CGM only to be reaccreted at a later time \cite[e.g.,][]{Marinacci10,Marinacci11,Ubler14,Armillotta16,DeFelippis17,Grand19,Huang22,Barbani23}. In more massive halos where feedback from AGN dominates, the evolution of the CGM will likely be different. AGN jets launch material at much higher velocities and are capable of quenching the galaxies they are in. We therefore expect that a Lagrangian selection of CGM gas in an AGN-dominated halo will spread out more in radius, with substantial fractions leaving the halo completely rather than adiabatically cooling within the halo as we consistently see in Figure \ref{f:galaxy_properties}. 

\subsection{Comparisons to Recent Work} \label{sec:comparisons}

There have been many recent efforts to study the properties and behavior of circumgalactic gas using different kinds of galaxy simulations, and we highlight some of these efforts here and connect them to our work. 

Using a set of idealized simulations of the CGM, \cite{Lochhaas20} measured the expected pressure support of gas in two different halo masses, one less massive and one more massive than the halos studied in this paper, and found a clear distinction between the two. While the larger halo ($\approx10^{12} \, M_{\odot}$) develops a hot phase and a cold phase, similar to what we find, the smaller halo ($\approx10^{11} \,M_{\odot}$) does not develop a multiphase CGM. Furthermore, the larger halo has a hot-phase-dominated CGM that is in hydrostatic equilibrium, indicating substantial pressure support, while the smaller halo lacks significant pressure support and is much more dependent on the precise form of feedback assumed in the model. Even without directly modeling any of the galaxy formation physics, \cite{Lochhaas20} showed that the environment and form of feedback impacting the CGM have huge implications for whether and how a multiphase distribution of gas forms and evolves and that our results may be different for less massive halos. 

\cite{Fielding20} analyzed the general properties of a varied set of galaxy simulations ranging from idealized to cosmological, allowing the effects of galaxies on the CGM to be quantified and compared to similar halo environments using different codes and under various assumptions regarding feedback. While the CGM in the cosmological simulations contained significantly more cool gas owing to cumulative accretion from the cosmic web, the global average thermal properties, particularly temperature and entropy, were remarkably consistent between the different models, especially for gas $\gtrsim 10^6 \, \rm{K}$. This suggests that the development of a volume-filling warm--hot, high-entropy phase like the one we find in our analysis should also arise in other types of CGM simulations of similar halos, although its relative importance compared to the cold phase may depend on the balance between cosmological accretion and energy injection from feedback, which would require further testing.

Another way to connect our results to other types of simulations is to compare timescales of multiple physical processes in the CGM, possibly including cooling, advection, freefall, and fragmentation. Many idealized simulations have found that ratios between these timescales determine crucial properties of gas in the halo, such as whether the hot phase remains adiabatic \citep[$t_{\rm{cooling}}/t_{\rm{advection}}$; e.g.,][]{Thompson16}, or whether it is stable to cooling or mixing in different environments \citep[$t_{\rm{cooling}}/t_{\rm{free-fall}}$; e.g.,][]{McCourt12,Sharma12,Prasad15,Voit17,Choudhury19}. In future work, we intend to examine the evolution of these timescales in detail and determine how they vary between idealized setups of galaxy--CGM systems and cosmological environments, which, as we previously discussed using \cite{Fielding20}, may affect the relative importance of multiple phases of gas over time.

Performing the kind of time-series analysis used in this study necessitates using cosmological initial conditions while also saving a large number of simulation snapshots. This is most often accomplished by running zoom-in simulations of individual halos rather than large cosmological boxes. \cite{Hafen20} used the \textsc{fire-2} simulations in this manner to determine where CGM gas ultimately ends up by following particle trajectories, and they found that their fates are very redshift dependent: $50\%$ of the $z=2$ gas in the CGM will move out of the halo on timescales $<1$ Gyr (i.e. around cosmic noon when stellar feedback is very efficient), but at lower redshifts ($z=0.25$) it takes about $3$ Gyr for the same fraction of the CGM to leave. This is a more significant mass exchange than occurs in our sample, where between $3\%$ and $11\%$ of the $z=1$ CGM is ever found within the central galaxy over a $2$ Gyr period (see Figure \ref{f:percgalaxy}), suggesting that the feedback in \textsc{fire-2} disrupts the CGM more effectively. Despite this, \cite{Hafen20} also found that gas, whether it is ultimately ejected from the halo completely into the IGM or ends up in the galaxy, often cycles into and out of different regions on the way. We see similar cycling at one of these interfaces in Figure \ref{f:percgalaxy}, though characterizing the full behavior would require further analysis around the halos' virial radii, where the resolution of our simulations is lower. Furthermore, a majority of the cold ($T < 10^{4.7} \, \rm{K}$) gas in those \textsc{fire-2} simulations ultimately accretes onto the central galaxy or a satellite galaxy in the halo by $z=0$, which is consistent with our picture of cold, \MgII--absorbing gas that forms at lower radii in the CGM and moves inward (seen in Figures \ref{f:T_dist} and \ref{f:ion_rdist}). 

Using the \textsc{foggie} simulations \citep{Peeples19}, which focus mass resolution on the diffuse CGM gas, \cite{Lochhaas23} find that the general assumption of hydrostatic equilibrium in the CGM is only roughly satisfied in the outer regions of halos averaged over long periods of time. Specifically, the inner CGM is always dominated by pressure from turbulence coming from feedback from the galaxy. At redshifts $z \gtrsim 0.5$, this turbulent pressure varies significantly with location and time, resulting in CGM properties that change with position in the halo on timescales $\lesssim 1$ Gyr, consistent with the variations in thermodynamic and kinematic quantities that we find in the subbox halos (Figures \ref{f:T_dist}, \ref{f:KPn_dist}, and \ref{f:autocorrelation}).

As discussed previously, our results are likely sensitive to the mode of feedback that is dominant in the galaxies, and thus the emergent properties of the evolving CGM gas could be different for other galaxies. \cite{Kelly22} examined this possibility by comparing the CGM of two different zoom-in simulation models -- \textsc{auriga} and \textsc{apostle} -- of otherwise similar Milky Way-mass galaxies. They are able to draw a key distinction between ejective and preventive feedback, which significantly affects the way the CGM develops. Ejective feedback, dominant in \textsc{auriga}, causes nearly all gas that feedback expels from the galaxy to remain in the CGM and eventually reaccrete at a later time, while in \textsc{apostle} most feedback is preventive and pushes nearly half the gas outside of the halo, as well as slowing accretion of pristine gas from the cosmic web. Our results using the CGM of TNG galaxies, which use a model very similar to the \textsc{auriga} simulations, unsurprisingly resemble the CGM of \textsc{auriga} galaxies, but the different CGM found in \textsc{apostle} suggests that a different subgrid physics model can result in a substantially different CGM, and a similar analysis done on those halos may yield very different results. Thus, fully determining how the CGM of different galaxies evolve will require determining which feedback models are more accurate at which redshifts. 

It is also possible that our results are somewhat sensitive to the spatial resolution of our simulations. \cite{Ramesh24} run a set of zoom-in simulations (GIBLE) with the TNG model to study this by using a refinement scheme based on \cite{Suresh19} that focuses on gas outside of the galaxy in the CGM. With this refinement, they are able to achieve a mass resolution in the CGM over $700\times$ better than the simulations studied here, as well as a subkiloparsec spatial resolution everywhere in the halo and under $100$ pc in the inner CGM. They find that many key global and time-averaged properties of the CGM, such as the total gas mass and temperature distribution, are converged at all of their resolutions, even though they consistently discover more cold gas structure at higher resolution in the form of clouds. This suggests that the two-phase CGM we identify would likely be persistent in simulations with better resolution. However, \cite{Ramesh24} also calculate CGM covering fractions for multiple ions and find that while hot gas traced by ions like \OVII{} is completely volume filling at all of their resolutions, the radial extent of covering fractions of both \MgII{} and \CIV{} increases at higher resolution. We therefore expect that in a higher-resolution simulation an emergent CGM cold phase like ours would extend to larger distances in the halo (beyond $\approx30-40$ kpc in Figure \ref{f:ion_rdist}), and \MgII--like gas in particular would be more common and longer-lasting, both due to the increased amount of cold structures that would regularly form throughout the halo.

\subsection{Connections to CGM Observations} \label{sec:observations}

The MEGAFLOW survey \citep{Bouche25} uses quasar absorption lines to study the cool gas in the CGM of star-forming galaxies around $z\approx1$ and has yielded many key results. \cite{Schroetter21} observed both \MgII{} and \CIV{} in the CGM of $z\approx1$ galaxies and found similar covering fractions as a function of impact parameter. However, by splitting populations in redshift bins, they found that the \MgII{} covering fraction is decreasing to lower redshift, while the \CIV{} covering fraction is only weakly increasing to lower redshift, potentially coevolving with the dark matter halo. \cite{Zabl19} used data from the same survey to show that \MgII{} in the CGM also displays a clear preference for corotation with galaxies at $z\approx1$, which many other works have also found using different modeling and simulation methods (\citealp[e.g., analytic model,][]{Sormani18}; \citealp[\textsc{pluto},][]{Sobacchi19}; \citealp[TNG/\textsc{arepo},][]{DeFelippis21}; \citealp[\textsc{gizmo},][]{Stern24}). There are also other observational surveys referenced in Section \ref{sec:intro} that have measured CGM covering fractions and kinematics, but we focus here on MEGAFLOW since it is specifically centered around $z\approx1$ and we defer a more comprehensive observational comparison to future work. 

With our analysis, we are able to potentially put results from MEGAFLOW into an evolutionary context. As shown in Figure \ref{f:ion_rdist}, the $z=1$ CGM only forms \MgII--like gas near the galaxy, and that gas is either infalling or rotating, indicating that existing corotating structures will persist for the next few Gyr. However, \MgII--like gas is regularly heated up and ejected outward into other phases (as well as cooling down and accreting inward) on timescales of $\approx500$ Myr, which suggests that gas corotating at any one time will likely be in a different temperature and kinematic state within $500$ Myr, and so the ``persistent'' rotating structure is constantly being disrupted and replenished. As the cold gas is heated to a \CIV--like phase, it generally expands to all radii over time, suggesting that the covering fraction of those warmer ions will increase as the halo grows over time, while the covering fraction of the cold \MgII--like phase that remains near the galaxy will not. 

Another key finding with observational implications is the differing mixing times associated with different ions. For example, as shown in Figure \ref{f:gasphasestability}, gas can remain \OVI--like for a few times longer than \CIV--like, despite otherwise evolving very similarly in radius. This complicates associations between different ions in the same CGM that have absorption lines consistent with the same velocity distribution. Due to the rapid and varied phase mixing timescales we see, we expect that observed phases with similar spectra may be probing gas in multiple different stages of evolution rather than a single coherent structure composed of multiple temperatures and densities.

\section{Summary} \label{sec:summary}

In this paper, we studied six halos with masses between $10^{11.5}$ and $10^{12} \, M_{\odot}$ from a subbox of the TNG100 cosmological simulation at a high time cadence to understand the behavior of CGM gas on timescales of $\sim1$ Gyr. We used the tracer particles of the TNG simulation to construct evolutionary paths of gas mass along multiple thermodynamic and kinematic quantities and then characterized the evolution of the $z=1$ CGM using populations of these tracer paths. Our conclusions are as follows:

\begin{enumerate}
    \item Over $50\%$ of the $z=1$ gas in the CGM remains in the halo for at least the following 2 Gyr; the rest of the CGM steadily cycles into and out of the galaxy through the galaxy--halo interface in the inner CGM, with generally only $2-8\%$ of the $z=1$ CGM being found within the galaxy at any one time (Figure \ref{f:percgalaxy}).

    \item Gas originating in the CGM tends to evolve within at most $\approx500$ Myr, regardless of its halo-centric radius at $z=1$, to a state that is characterized by a cold ($T\approx10^4 \, \rm{K}$), low-entropy phase closer to the galaxy and a warm--hot ($T\approx10^{5.5} \, \rm{K}$), high-entropy phase farther from the galaxy. Transfer between these phases occurs when stellar feedback heats and ejects cold gas that is near or has joined the galaxy at large radial velocity and when gas slowly loses angular momentum and moves inward while cooling (Figures \ref{f:T_dist} and \ref{f:KPn_dist}). 

    \item In conjunction with this temperature evolution, gas at lower pressures and densities moves to larger radius over time, whereas gas at higher pressures and densities remains close to the galaxy (Figure \ref{f:KPn_dist}).

    \item Over a $0.5$ dex range in halo mass, the thermal evolution of CGM gas is very consistent, despite as much as an order-of-magnitude difference in the galaxy mass and SFR. Consistently larger SFRs in the galaxy are likely correlated with a more massive CGM at all temperatures (Figure \ref{f:galaxy_properties}). 

    \item By computing autocorrelations of angular momentum and radial velocity, we find that cold CGM gas nearest the galaxy ``forgets'' its kinematic history more quickly than the warm--hot phase does, indicating that feedback disrupts and reprocesses cold gas in the inner CGM to a greater extent than warm--hot gas (Figures \ref{f:deltar_vr} and \ref{f:autocorrelation}). 

    \item CGM gas associated with commonly observed ions at a single redshift transitions out of being associated with those ions and mixes with others. The speed of this mixing varies and is strongly ion dependent: a majority of the \OVI--like gas remains \OVI--like for $250-750$ Myr, while most of the \CIV--like and \MgII--like gas transitions away from its phase after $50-250$ Myr (Figures \ref{f:gasphasefrac} and \ref{f:gasphasestability}). 

    \item \MgII--like gas always appears and remains near the galaxy, as it is composed primarily of the cold low-entropy phase identified before, while both \CIV--like and \OVI--like gas, which trace higher temperatures and lower densities, spread out throughout the entire halo, regardless of the spatial origin of the gas (Figure \ref{f:ion_rdist}).

    \item Entropy radial profiles of the warm--hot phase are flat within the CGM for all halos in our sample, while the entropy profiles of the cold phase vary by as much as $1$ dex between different halos (Figure \ref{f:Kprofile}). This, along with the correlation we see between entropy and radial velocity (Figure \ref{f:KPn_dist}), encourages us to think of galactic feedback as a convective-like process in which feedback-heated fluid elements rise while radiatively cooling ones fall.
\end{enumerate}

With this work, we have demonstrated how multiple physical processes occurring simultaneously, likely including gas mixing and feedback from supernovae, can affect the way in which gas in the CGM evolves. Despite the complexity of the environment and potential merger histories, we predict that a two-phase CGM will emerge following the mixing of multiple states of gas at $z=1$ for galaxy--halo systems in which ejective supernova feedback is dominant, though the extent and kinematic behavior of those two phases, particularly the cold phase, may be dependent on the spatial resolution. In future studies, this kind of analysis can be used to establish how sensitive our conclusions are to redshift, environment, feedback model, and gas morphology and resolution (by, e.g., modeling distributions of cold clouds with tools like \textsc{cloudflex} \citep{Hummels24}), and whether there is a common theoretical expectation for how the CGM and galaxy coevolve.

\begin{acknowledgments}

D.D. thanks Dr. Adelle Bergman for the inspiration, encouragement, and unwavering love and support that allowed this and many other works to come to fruition. We thank the anonymous referee for their detailed and insightful comments. D.D. acknowledges support from internal University of Arizona funds provided by the Office of Research and Partnerships. The Flatiron Institute is supported by the Simons Foundation. G.L.B. acknowledges support from the NSF (AST-2108470, AST-2307419), NASA TCAN award 80NSSC21K1053, and the Simons Foundation through the Learning the Universe Collaboration.

\end{acknowledgments}

\vspace{5mm}

\software{\textsc{NumPy} \citep{vanderWalt11}, 
          \textsc{Matplotlib} \citep{Hunter07}, 
          and \textsc{IPython} \citep{Perez07}.
          }

\appendix
\section{Behavior of individual tracers} \label{sec:appendix}

\begin{figure*}
\fig{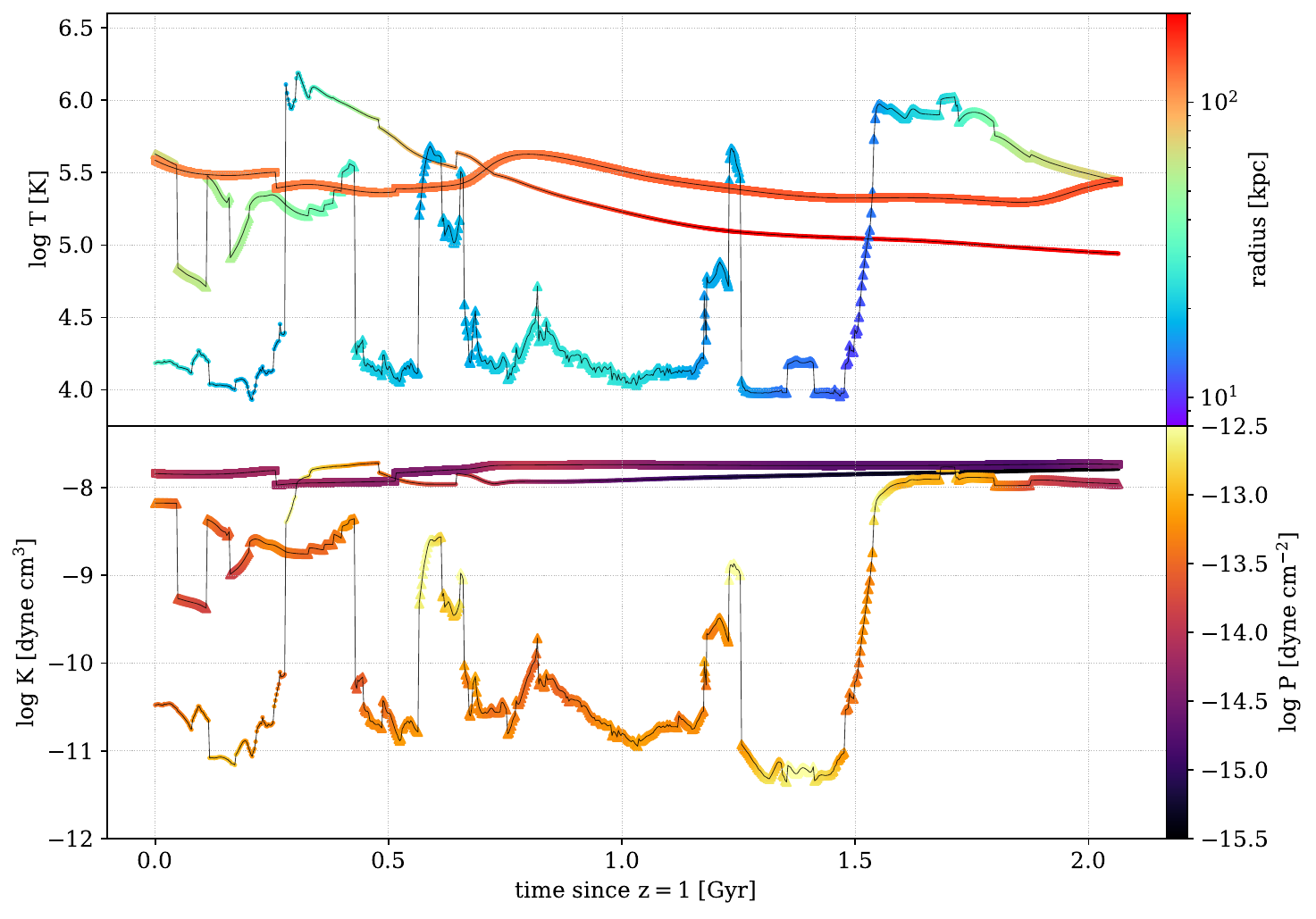}{0.99\textwidth}{}
\vspace{-15pt}
\caption{Temperature (top) and entropy (bottom) evolution of three tracers from halo5819 colored by their radius and pressure, demonstrating the variability in the behavior of individual tracers compared to the population studied in the rest of the paper.
}
\label{f:individual_tracers}
\end{figure*}

In this appendix, we describe how individual gas tracers used in our analysis evolve over time compared to the evolution of the populations of tracers considered throughout the rest of the paper. In Figure \ref{f:individual_tracers}, we plot the temperature and entropy evolution of three tracers from the same halo as Figures \ref{f:T_dist} and \ref{f:KPn_dist}, starting at three different positions in the halo. Here we see three different possible paths tracers can take in the CGM: they can remain unaffected by any feedback from the galaxy and stay at a relatively constant temperature and radius, they can be heated by the galactic wind close to the galaxy and then spend the next Gyr moving outward and slowly cooling (a behavior associated with a wind as described in \citealp{Chevalier85}), and they can be heated by winds multiple times, with only the last time resulting in a long-term change in phase and radius. We also plot the entropy and pressure of the same three tracers as a function of time. With these quantities, we show that the tracers that slowly cool from $T=10^6 \, \rm{K}$ are cooling adiabatically and that tracers experience higher pressures when they are at moderate or lower entropy, or when they are nearer to the galaxy and more affected by feedback. 

\bibliography{references}{}
\bibliographystyle{aasjournal}

\end{document}